\documentclass{aa}  
\usepackage{graphicx}
\usepackage{txfonts}
\usepackage{xcolor}
\usepackage{orcidlink}
\usepackage{ulem}
\usepackage{hyperref} 
\def\apj {ApJ}
\def\apjl {ApJL}
\def\apjs {ApJS}
\def\aj {AJ}
\def\aap {A\&A}
\def\mnras {MNRAS}

\def\nat {Nature}

\def\R200 {R_{200}}

\def\Gyr {\rm Gyr}

\begin{document} 

\authorrunning{S. Levis et al.}

   \title{Galaxy evolution in groups: Transition galaxies in the IllustrisTNG simulations}

  \author{Selene Levis\inst{1,2}\orcidlink{0000-0003-1887-776X},
   Valeria Coenda\inst{1,3}\orcidlink{0000-0001-5262-3822},
   Hern\'an Muriel\inst{1,3}\orcidlink{0000-0002-7305-9500},
   Mart\'in de los R\'ios\inst{4,5,6}\orcidlink{0000-0003-2190-2196},
   Cinthia Ragone-Figueroa \inst{1,7,8}\orcidlink{0000-0003-2826-4799},
   H\'ector J. Mart\'inez\inst{1,3}\orcidlink{0000-0003-0477-5412}
   \& Andr\'es N. Ruiz \inst{1,3}\orcidlink{0000-0001-5035-4913}
}

\institute{
        Instituto de Astronom\'ia Te\'orica y Experimental, CONICET - UNC, Laprida 854, X5000BGR, C\'ordoba, Argentina
        \and
        Facultad de Matem\'atica, Astronom\'ia, F\'isica y Computaci\'on, Universidad Nacional de C\'ordoba, Av. Medina Allende s/n, X5000HUA, C\'ordoba, Argentina
        \and
        Observatorio Astron\'omico, Universidad Nacional de C\'ordoba, Laprida 854, X5000BGR, C\'ordoba, Argentina
        \and
        Departamento de F\'isica Te\'orica, Universidad Aut\'onoma de Madrid, Cantoblanco, 28049 Madrid, Espa\~na 
        \and
        Instituto de F\'isica Te\'orica (IFT-UAM/CSIC), Universidad Aut\'onoma de Madrid, 28049 Madrid, Espa\~na
        \and
        SISSA - International School for Advanced Studies, Via Bonomea 265, 34136 Trieste, Italy
        \and
        INAF, Osservatorio Astronomico di Trieste, via Tiepolo 11, I-34131, Trieste, Italy
        \and
        IFPU, Institute for Fundamental Physics of the Universe, via Beirut 2, 34014 Trieste, Italy
}

   \date{Received XXXX; accepted XXXX}

%%%%%%%%%%%%%%%%%%%%%%%%%%%%%%%%%%%%%%%%
 
   \abstract
   % CONTEXT
   {The evolution of galaxies is significantly influenced by the environments they inhabit. While high-density regions, such as clusters of galaxies have been widely studied, the dynamics and quenching processes in intermediate environments remain less explored. These systems provide a valuable context for understanding the transition of galaxies from active star formation to quiescence.} 
   % AIMS 
   {This study aims to characterise the astrophysical properties of galaxies in intermediate-mass galaxy groups ($13.5 \leq \log(M_{200}/M_{\odot}) \leq 13.7$), with a focus on their evolutionary pathways and the key processes driving their transition through the green valley (GV) and green zone (GZ). Specifically, we explore the interplay between internal and external quenching mechanisms and their impact on galaxy evolution within groups and their surrounding environments.} 
   % METHODS
   {Using the Illustris TNG300-1 hydrodynamical cosmological simulations, we classified galaxies based on their trajectories and environment into five categories: group galaxies (GRs), backsplash galaxies (BSs), recent infallers (RINs), infall galaxies (INs), and field galaxies (FGs). We examined their optical colours in the $(u-r)$-stellar mass diagram, specific star formation rates (sSFRs), gas fractions, and stellar mass evolution from $z=0.5$ to $z=0$.} 
   % RESULTS
   {At $z=0$, FGs dominate the blue cloud, while GRs show progressive reddening, especially in low-mass systems. Compared to the other classes, BSs exhibit the highest fraction of green galaxies, highlighting their transitional nature. On the other hand, RINs show a rapid quenching upon entering $R_{200}$. Notably, RINs  experience greater environmental influence than BSs, due to their later entry into more massive systems.}
   % CONCLUSIONS
   {Our results reveal that the timing of group entry and environmental effects, such as gas depletion, are critical to the decline in sSFR and the transition of galaxies through the GV towards quiescence. Green BS and RINs, in particular, demonstrate distinct evolutionary tracks shaped by their interactions with the group environment, with green RINs  showing more rapid changes over shorter timescales. This analysis highlights the significant role of the entry time into the group in shaping galaxy evolution. BSs, having entered at an earlier stage, encounter a system that is less massive than the one RINs interact with upon arrival. Consequently, RINs experience a stronger influence from the intragroup medium than BSs do.}

   \keywords{
   Galaxies: general --
   Galaxies: groups: general --
   Galaxies: evolution --
   Galaxies: star formation --
   Galaxies: kinematics and dynamics
   }

\maketitle

%%%%%%%%%%%%%%%%%%%%%%%%%%%%%%%%%%%%%%%%

\section{Introduction}\label{sec:intro}

The observed range of the astrophysical properties of galaxies, such as their mass, size, morphology, colour, and star formation rate is a consequence of both secular processes with INs and the action of the environments the galaxy has inhabited. Galaxies in the nearby Universe can be broadly classified into two main groups: 1) red galaxies, characterised by ageing stellar populations and predominantly found in high-density regions, and 2) blue, star-forming galaxies, which are preferentially located in low-density regions. This distinctive bimodal distribution is evident in the optical colour-magnitude diagram (CMD, or colour-stellar mass), as demonstrated in a number of previous studies (e.g. \citealt{Strateva:2001, Kauff:2003, Baldry:2006}).

Within the CMD, star-forming galaxies are commonly referred to as the 'blue cloud' (BC), whereas passive galaxies constitute the red sequence (RS). The intermediate region between these two populations is widely recognised as the green valley (GV; \citealt{Wyder:2007}). The GV is typically considered a transitional region between the BC and RS (e.g. \citealt{Schawinski:2007a, Mendez:2011}), which host galaxies that have recently experienced a decline in their star formation activity (e.g. \citealt{Salim:2007, Salim:2014, Smethurst:2015, Coenda:2018, Phillipps:2019, Trussler:2020, Sampaio:2022}). On the other hand, some passive galaxies may undergo a phase of rejuvenation, when a new burst of star formation occurs following a period of reduced activity. This process can shift galaxies from the RS back into the GV and even further down into the BC (e.g. \citealt{Graham:2017, Rowlands:2018, Nelson:2018, Chauke:2019, Parente:2024}). In this context, studying the properties of the GV galaxies as a function of the stellar mass and environment could provide insights into the physical mechanisms behind the suppression of their star formation.

Various selection criteria have been proposed to define the GV. Traditional methods often rely on colour to delineate the GV, including  $NUV - r$ \citep{Wyder:2007, Lee:2015, Coenda:2018}, $U - B$ \citep{Mendez:2011}, $g - r$ \citep{Trayford:2015, Trayford:2016, Walker:2013, Eales:2018}, and $u - r$ \citep{Schawinski:2014, Phillipps:2019, Parente:2024}. Other authors have employed the specific star formation rate \citep[sSFR;][]{Schiminovich:2007, Salim:2014, Phillipps:2019, Starkenburg:2019}, the star formation rate \citep[SFR;][]{Noeske:2007, Chang:2015}, or the $4000 \AA$ break strength \citep{Angthopo:2019}. \citet{Nyiransengiyumva:2021} found that selecting GV galaxies based on UV and optical data encompasses a wide variety of galaxy types in terms of stellar mass, sSFR, morphological classification, spectroscopic type, and other parameters, revealing no particular bias towards either star-forming or passive galaxies. In contrast, they observed that defining the GV in terms of sSFR introduces a bias favouring star-forming galaxies. More recently, \citet{Pandey:2024} introduced a novel definition of the GV using entropic thresholding on the colour-stellar mass plane, offering a natural boundary for the GV. In addition, \citet{Parente:2024} modelled the $(u - r)$ colour distribution using Gaussian mixture models (GMMs), assessing whether the population is better represented by a single or bimodal distribution. This statistical approach provides an alternative method to identify GV galaxies based on the underlying colour distribution.

Numerical simulations, such as NIHAO \citep{Wang:2015},  Illustris TNG \citep{Nelson:2019}, and EAGLE (\citealt{Schaye:2015, Crain:2015}), have become essential tools for studying the transition of galaxies into the GV. Several authors have identified active galactic nucleus (AGN) feedback as the dominant mechanism driving this transition (e.g. \citealt{Trayford:2016, Blank:2022}). These simulations also shed light on the quenching timescales and their connection to galaxy properties such as stellar mass, as well as to environmental factors \citep{Nelson:2018, Wright:2019}.

Numerous mechanisms have been proposed as potential drivers of quenching INs. These mechanisms are commonly categorised into two groups: internal mechanisms, which are influenced by the characteristics of the galaxy itself, and external mechanisms that depend on the environment in which the galaxy resides. Different quenching mechanisms operate on different spatial and temporal scales and are responsive to specific structural components withINs.

Internal mechanisms are predominantly associated with a galaxy's stellar mass, and are collectively known as mass quenching. Several proposed physical processes fall under this category, including supernova-driven winds (e.g. \citealt{Bower:2012, Stringer:2012}), halo heating \citep{Marasco:2012}, feedback from massive stars (e.g. \citealt{DallaVecchia:2008, Hopkins:2012}), and AGN feedback (e.g. \citealt{Nandra:2007, Hasinger:2008, Silverman:2008, Cimatti:2013}). 
On the other hand, external quenching is a consequence of environmental factors. While the impact of the environment has been probed in galaxy clusters (e.g. \citealt{Coenda:2018,GonzalezDelgado:2022,Chang:2022}), the suppression of star formation driven by environmental factors also takes place in less dense groups (\citealt{Barsanti:2018, Davis:2019}).
When galaxies traverse the inner regions of clusters or groups of galaxies, they undergo various processes, including tidal stripping (e.g. \citealt{Zwicky:1951, Gnedin:2003a, Villalobos:2014}), thermal evaporation \citep{Cowie:1977}, and interactions or harassment among galaxies (e.g. \citealt{Moore:1996, Moore:1999, Gnedin:2003b}). 

In galaxy groups, the likelihood of mergers is higher compared to galaxy clusters, providing a mechanism that promotes the formation of early-type galaxies. Ram-pressure stripping of cold gas is a well-documented phenomenon in clusters of galaxies, supported by such studies as as \citet{G&G:1972}, \citet{Abadi:1999}, \citet{Book:2010}, and \citet{Steinhauser:2016}, to name just a few. It can also occur in less massive systems such as galaxy groups, as evidenced in the works by \citet{Rasmussen:2006}, \citet{Jaffe:2012}, and \citet{Hess:2013}.
Interactions between galaxies and the intra-group gas, such as strangulation, can lead to the removal of warm and hot gas from a galactic halo, thus disrupting the gas supply needed for star formation (e.g. \citealt{Larson:1980, Kawata:2008}). According to \citet{Peng:2015}, strangulation has been proposed as the primary mechanism responsible for quenching star formation in local galaxies, operating on typical timescales of $\sim 4\ \Gyr$. Additionally, \citet{Kawata:2008} have predicted that strangulation also plays a role in galaxy groups.

Throughout their life cycles, galaxies encounter diverse environmental conditions and can be influenced by one or more mechanisms at various stages. As they fall towards larger systems, the physical processes they experience depend on their origin: whether they are part of a group (e.g. \citealt{McGee:2009, deLucia:2012, Wetzel:2013}; \citealt{Hou:2014}), falling in from the field (e.g. \citealt{Berrier:2009}), or travelling through filamentary streams (e.g. \citealt{Colberg:1999, Ebeling:2004, Martinez:2016, Salerno:2019}; \citealt{Rost:2020, Kuchner:2022}). Prior to entering a system, galaxies may undergo physical transformations driven by these interactions in a process collectively referred to as pre-processing \citep[e.g.][]{Mihos:2004, Fujita:2004}.

After being incorporated into a cluster or group, galaxies may either remain gravitationally bound to the system or be ejected to distances exceeding several $R_{200}$, where $R_{200}$ denotes the radius encompassing 200 times the mean density of the Universe (e.g. \citealt{Mamon:2004, Gill:2005, Rines:2006, Aguerri:2010, Muriel:2014, Casey:2023, Ruiz:2023}). These galaxies could turn around and return to the cluster during subsequent infall events at a later time. This distinct population, referred to as backsplash (BS) galaxies \citep{Balogh:2000}, serves as a valuable resource for investigating environmental effects on galaxy properties and identifying the most critical stages of their evolution. In recent years, many theoretical studies have focussed on the investigation of BSs in clusters of galaxies (e g. \citealt{Haggar:2020, Knebe:2020, Benavides:2021, Borrow:2023, Hough:2023, Ruiz:2023}).

In this paper, we use state-of-the-art cosmological simulations to study the evolution of transition galaxies in and around low-mass systems of galaxies with $13.5 \leq \log(M_{200}/M_{\odot}) \leq 13.7$. We specifically examine the evolution of various astrophysical properties of backsplash galaxies and those that have recently fallen into groups. These galaxies have undergone similar quenching mechanisms, albeit at different times and, arguably, with varying intensities. This difference can be attributed to the fact that the former have interacted with the virial region of groups at a less evolved stage compared to the latter. The article is structured as follows. In Sect. \ref{sec:sample}, we provide a detailed description of the simulated galaxy catalogue and define the galaxy types considered. Section \ref{sec:gv} outlines the criteria used to identify transition galaxies. In Sect. \ref{sec:comparison}, we examine the evolution of galaxies in the colour-magnitude diagram as a function of their environment. Section \ref{sec:green} focusses on comparing the astrophysical properties of the GV galaxies in our sample. Finally, in Sect. \ref{sec:conclusions}, we summarise the conclusions of our analysis.

%%%%%%%%%%%%%%%%%%%%%%%%%%%%%%%%%%%%%%%%

\section{The sample}\label{sec:sample}  

For our analysis, we constructed samples of groups and galaxies drawn from the TNG300-1 gravo-magnetohydrodynamical simulations \citep{Nelson:2019}. These simulations are based on a $\Lambda$CDM cosmological model, with parameters drawn from \citet{PlanckColab:2016}, namely: $\Omega_{\Lambda,0}=0.6911$, $\Omega_{m,0}=0.3089$, $\Omega_{b,0}=0.0486$, $\sigma_{8}=0.8159$, $n_{s}=0.9667$, and $h=0.6774$. The simulations were conducted using the moving-mesh code \textsc{Arepo} developed by \citet{Springel:2010}.

The IllustrisTNG\footnote{https://www.tng-project.org/} project encompasses three main simulation volumes: TNG50, TNG100, and TNG300. Each volume incorporates a sophisticated model for galaxy formation physics, allowing for the accurate simulation of galaxy evolution from a redshift of $z=127$ to $z=0$. In particular, TNG300 offers a high simulation volume that is ideal for studying galaxy clustering and analysing rare, massive objects such as galaxy clusters and groups, providing a statistically substantial sample of galaxies.
The IllustrisTNG project provides 100 snapshots of its cosmological volumes. For each snapshot, a friends-of-friends (FoF) halo catalogue and a \textsc{Subfind} \citep{Springel:2001} sub-halo catalogue were delivered.

Throughout this work, we restricted our analysis to galaxies with $\log(M_{\star}/M_{\odot}) \geq 9.5$ at $z=0$ from the Illustris TNG300-1 simulation. This minimum stellar mass threshold was set to exclude the excess of low-mass red galaxies produced by the simulation, which are predominantly satellites in dense environments. For further details, we refer to \citet{Nelson:2018}.

\subsection{Galaxies in groups and in their surroundings}

We selected a total of 214 groups from the FoF group catalogue, with masses in the range of $13.5 \leq \log(M_{200}/M_{\odot}) \leq 13.7$, where $M_{200}$ is the mass contained within the region of radius $R_{200}$ that encompasses 200 times the critical density. This upper mass limit is consistent with \citet{Paul:2017}, who highlight significant differences in the physical properties of galaxy clusters and groups, identifying the transition around $\log(M_{200}/M_{\odot}) \sim 13.9$. The chosen lower limit ensures that all selected systems contain more than five galaxies. The selected mass range corresponds to intermediate-mass systems, where environmental effects are less extreme, compared to their higher-mass counterparts. By focussing on groups, where environmental influences are more moderate, we seek to better understand how the dynamical histories of galaxies in these systems impact their star formation trajectories and, more specifically, how and when galaxies transition into the GV. Our sample of galaxies associated with these groups comprises 5103 galaxies.

Following \citet{delosRios:2021}, we selected all halos at $z=0$ that have no companion halos more massive than $0.1 \times M_{200}$ within a distance of $5 \times R_{200}$. This selection is aimed at removing systems undergoing major mergers or interacting with massive companions, which could affect the orbits of galaxies near the groups. We classified all the galaxies into different classes based on their 3D orbits around galaxy groups, following the definitions given in \citet[as detailed in their Fig. 1]{delosRios:2021}, namely:

\begin{enumerate}

\item Group galaxies (GRs): Galaxies that first crossed $R_{200}$ more than 2 Gyr ago. If they have exited $R_{200}$ at least once, they must have re-entered this radius on at least two occasions. Most of them are located within $R_{200}$, although some may temporarily be found outside this radius in their orbital motion. This sample consists of $1452$ galaxies. On average, $6.8$ GRs are associated with each group.

\item  Backsplash galaxies (BSs): These galaxies have crossed $R_{200}$ twice: once upon entering the group and once again when leaving. Currently situated outside $R_{200}$, they have only traversed the group once and are potential candidates to become group galaxies in the future. In our sample, $486$ galaxies have crossed $R_{200}$ twice. The average number of BSs  per group is $2.3$.

\item Recent infallers (RINs): Found within $R_{200}$, these galaxies have crossed this boundary only once in the last $2\,\Gyr$ ago. They are experiencing the environmental impacts of the group for the first time and may evolve into BSs in the future. This sample comprises $333$ galaxies. Our groups host   $1.6$ RINs,  on average.

\item Infall galaxies (INs): These galaxies have spent their entire existence outside of $R_{200}$  and are now in the process of falling towards the group, as indicated by their negative radial velocities relative to the group. This sample consists of $2832$ galaxies, which results in a mean of 13.2 INs  per group. 

\end{enumerate}

\citet{delosRios:2021} considered a fifth type of galaxy, defined based on its trajectory and referred to as interlopers (ITLs). These galaxies have no physical connection to the system, have never approached the group centre within $R_{200}$, and are located in halos that are receding from the group. 
In this work, we decided to exclude ITL galaxies because they are often associated with FoF halos that are more massive than our groups. Consequently, ITLs exhibit signs of pre-processing, making them unsuitable to represent FGs.

\subsection{Field galaxies}

As a comparison sample, we randomly selected $35,000$ galaxies from the simulation box, regardless of their environment. These galaxies can reside in both high-density and low-density regions within the box. This random sample is used exclusively to define the GV and GZ in the next section.

From this random sample, we selected a sub-sample of galaxies located outside systems. FGs are defined as those located more than $3 R_{200}$ away from the centres of all FoF groups, with $\log(M_{200}/M_{\odot}) \geq 13$ at $z=0$. The resulting sample comprises $25,390$ FGs. This galaxy sample is used to compare the properties of galaxies in groups with those in the field.

%%%%%%%%%%%%%%%%%%%%%%%%%%%%%%%%%%%%%%%%

\section{Green galaxies definition} \label{sec:gv} 

\begin{figure}
\centering
{\includegraphics[width=0.43\textwidth]{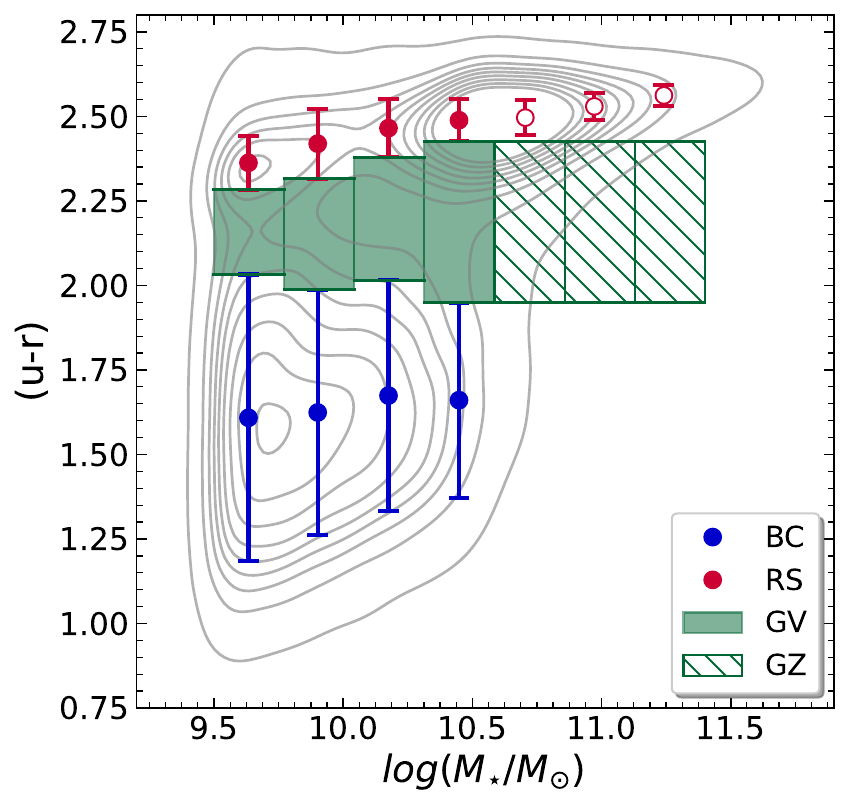}
\caption{\label{fig1}  $(u-r)$ colour vs stellar mass diagram (CMD) for our random sample independent of the environment. The CMD density isocontours are shown in gray. Points with error bars represent the mean and standard deviation of Gaussian fits for the blue cloud (BC, blue symbols) and the red sequence (RS, red symbols) as a function of stellar mass. The shaded region represents the   GV, which comprises $13.15\%$ of the sample of galaxies, while the hatched area designates the  GZ, which contains $3.75\%$ of the sample. 
}}
\end{figure}

We show in Fig. \ref{fig1} the distribution of $(u-r)$ colour versus stellar mass for our environment-independent random sample. We resorted to a Gaussian fitting approach to distinguish the blue and red populations (e.g. \citealt{Baldry:2004, Martinez:2006, Zandivarez:2011, Smethurst:2015, Vulcani:2015, Coenda:2018, Parente:2024}).

We divided the galaxy sample into seven stellar mass bins, ranging from $\log(M_{\star}/M_{\odot}) = 9.5$ to $11.5$, and we analysed the $(u-r)$ colour distribution within each mass range. Firstly, we determined whether each mass bin shows a single population or if it is better represented by two populations: blue and red. As in \citet{Parente:2024}, we used the Bayesian information criterion (BIC) to guide this decision, except in the fourth mass interval (for reasons explained below). The BIC favours models that achieve a good fit using few parameters, balancing model complexity with fit quality. The Gaussian models used to calculate the BIC were fitted using Python’s \textsc{GaussianMixture} class. Additionally, we ensured that no Gaussian component had a weight below 0.2 and that the areas within one standard deviation of the two Gaussians did not overlap. When these conditions were met, we selected a two-Gaussian model; otherwise, we used a single Gaussian model.

For the first three mass ranges, the best fit was achieved with a two-Gaussian model representing the blue and red populations, which were fitted using Python’s \textsc{GaussianModel} class. For the last three mass bins, a single Gaussian was used to fit the red population, as the blue population was no longer present. In these cases, we identified the maximum peak of the distribution and selected the values from this peak to the nearest tail. We then reflected these values to create a symmetric distribution, providing a more accurate representation of the red population.

In the fourth mass range, we observed the characteristic bimodal distribution, along with the emergence of a third population. Similarly to the approach in \citet{Vulcani:2015}, we obtained the best fit here by applying a three-Gaussian model, implemented with Python’s \textsc{CurveFit} class. 

We identified two transition regions in Fig. \ref{fig1}: the GV itself, where both blue and red populations coexist, and the green zone (GZ), where only the red population persists. The boundaries of these regions are defined by the central values and standard deviations of the Gaussian fits within each mass bin.

The GV was defined for each stellar mass range using fixed values in colour, where the upper boundary is given by $\mu_R - \sigma_R$ and the lower boundary by $\mu_B + \sigma_B$, where the sub-scripts $B$ and $R$ refer to the blue and red Gaussians fits, respectively. The GV is well defined in the first four mass bins.
For higher mass bins, the colour distribution is better fit by a single Gaussian function representing the red population. For these bins, we defined the GZ limits using the same limits as the GV of the fourth mass bin. 
In this study, we classified all galaxies that fall within these regions as 'green galaxies'. Those above this range were classified as 'red galaxies', and those below were classified as 'blue galaxies'.

%%%%%%%%%%%%%%%%%%%%%%%%%%%%%%%%%%%%%%%%

\section{Comparing red, green, and blue galaxies by class} \label{sec:comparison}

\subsection{Colour-mass diagram at $z=0$} \label{sec:cmd_z0}

\begin{figure*}
\centering
{\includegraphics[width=1\textwidth]{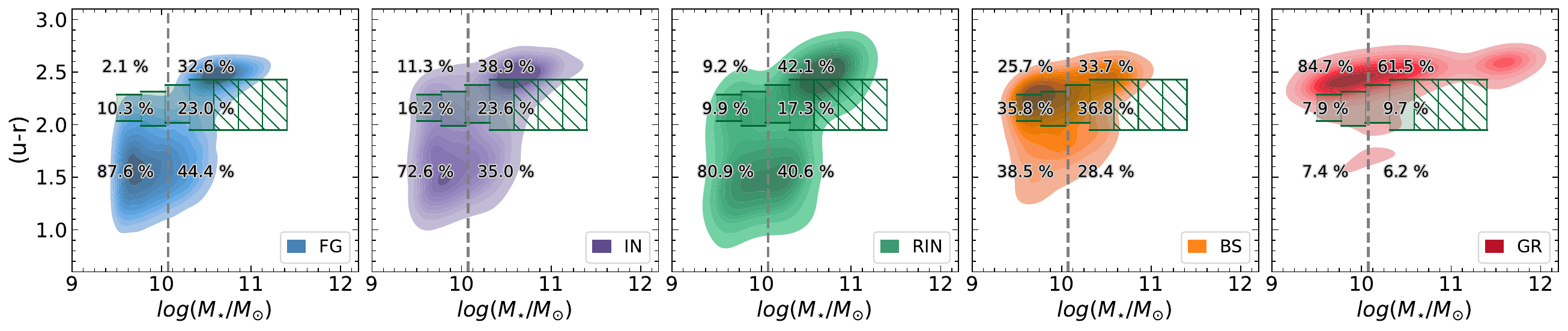}
\caption{\label{fig2}  $(u-r)$ colour-stellar mass diagram at $z=0$ for each galaxy class. From left to right, the panels correspond to FGs, INs, RINs, BSs, and GRs, respectively. In these panels, different colours correspond to different galaxy number density, ranging from the lowest (lighter colours) to the highest (darker colours) values in each case. In each plot, the shaded region indicates the GV, while the hatched area marks the GZ. The dashed gray vertical line denotes the median $\log(M_{\star}/M_{\odot})= 10.07$  for our random sample independent of the environment, delimiting the low- and high-mass galaxies. The percentages indicated on the plots represent the fractions of blue (bottom), green (middle), and red (top) galaxies in the low and high stellar mass sub-sample.
}}
\end{figure*}

In this section, we analyse the astrophysical properties of galaxies in our samples as defined in Sect. \ref{sec:sample}. Our goal is to understand how galaxies in groups and their surroundings evolve from star-forming to quiescent states. Additionally, we compare these galaxy properties with those observed in FGs to gain insight into the influence of environmental factors on galactic evolution.

Figure \ref{fig2} shows the $(u-r)$ colour-stellar mass diagram for each galaxy class at $z=0$. Contours indicate the galaxy number density. The boundaries of the GV and the GZ are indicated by the shaded and hatched regions, respectively. The median mass of our random sample independent of the environment is $\log(M_{\star}/M_{\odot})= 10.07$, which defines the threshold between what hereafter we  refer to hereafter as the low- and high- mass galaxies.

We observe that FGs predominantly reside in the BC. An evolutionary trend is apparent for galaxies associated with galaxy systems, where galaxies gradually redden the longer they remain within the system, that is, when observing the classes (apart from the FGs) from left to right. This trend is particularly evident in low-mass galaxies, which tend to be satellites that are strongly affected by environmental processes.

Most INs and RINs are located in the BC, indicating a prevalence of star-forming characteristics. Additionally, INs exhibit similar proportions of blue, green, and red populations than RINs. In the transitional phase, BSs show the highest fraction of green galaxies. On the other hand, GRs are predominantly found in the RS, with only a few blue or green galaxies. This pattern highlights the role of environmental interactions in driving galaxy evolution, shifting galaxies towards quiescence as they spend more time in denser regions.

When considering the fraction of green galaxies respect to the total sample, we observe a systematically lower fraction of low-mass galaxies compared to high-mass galaxies, except for BSs. BSs not only display the highest fraction of green galaxies, but also show no clear dependence on stellar mass.

Another observed trend is the low percentage of low-mass red galaxies among FGs. In contrast, galaxies around groups exhibit a gradual increase in this percentage, progressing from IN to GRs. The similar fraction of red galaxies observed between IN and RINs may reflect pre-processing effects experienced by these galaxies as they approach denser regions.

\subsection{Evolution of the colour-mass diagram} \label{sec:evolution}

\begin{figure*}
\centering
{\includegraphics[width=0.95\textwidth]{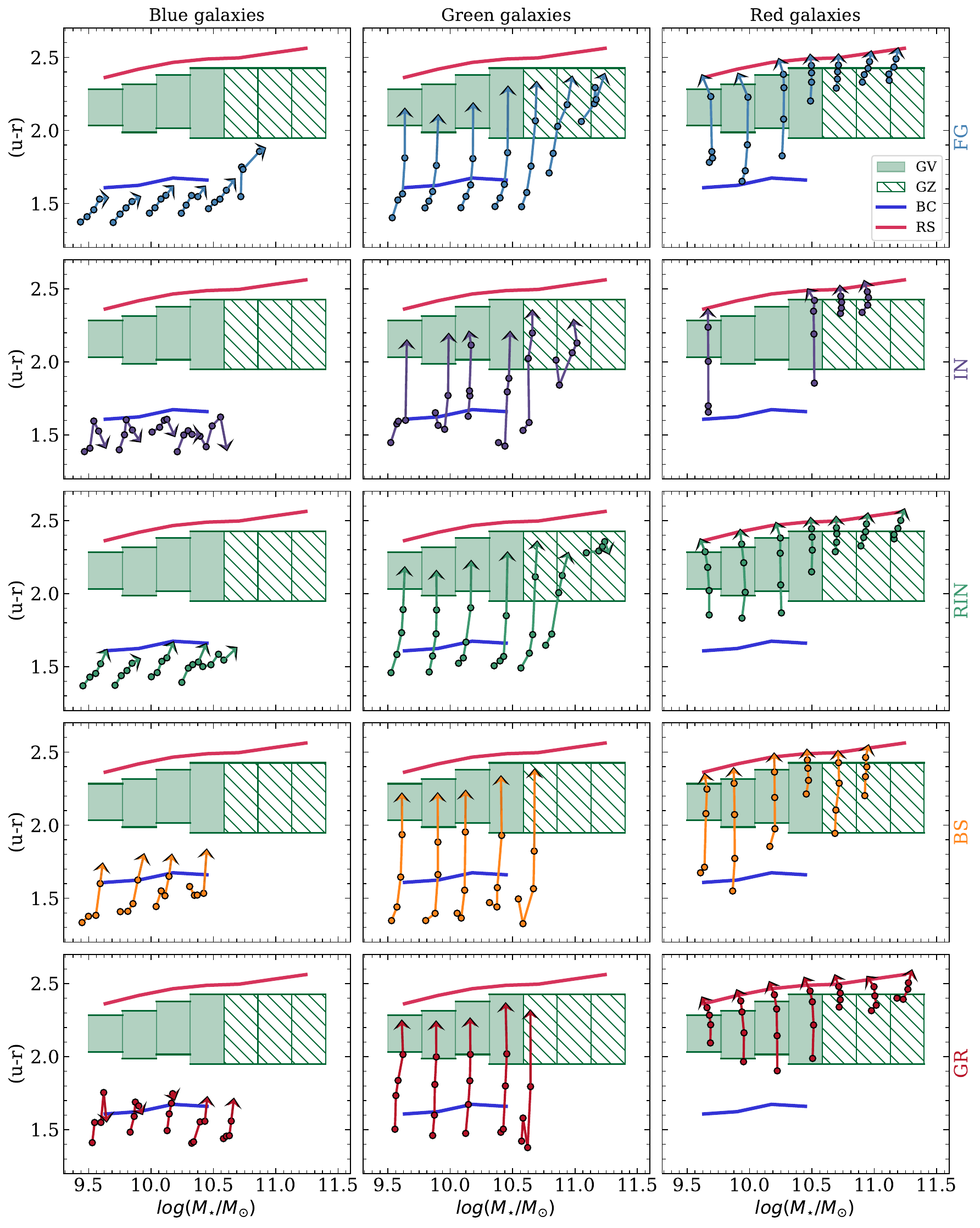}
\caption{\label{fig3} Evolutionary tracks in the $(u-r)$ colour versus stellar mass diagram for galaxies classified as blue (left column), green (middle column), and red 
(right column) at $z=0$. The position of the arrow tips indicates the median location of galaxies at $z=0$ for each stellar mass bin used to define the GV and the GZ in Sect. \ref{sec:gv}. The circles track the median positions of galaxies in the diagram at previous redshifts $z=0.1$ ($t=1.34$ Gyr), $z=0.2$ ($t=2.48$ Gyr), $z=0.3$ ($t=3.50$ Gyr) and $z=0.4$ ($t=4.41$ Gyr) for each bin. The shaded region represents the GV, while the hatched area marks the GZ. The blue and red lines represent the mean colour of BC and RS, respectively. Results are presented for FGs (first row), INs (second row), RINs (third row), BSs (fourth row), and GRs (fifth row).
}}
\end{figure*}

To understand the origins of galaxies classified as blue, green, or red at $z=0$, we analyse in Fig. \ref{fig3} their median evolutionary trajectories in the CMD. Using the stellar mass bins that define the GV and GZ, we track the median positions of galaxies backwards from $z=0$  to earlier epochs, specifically at  $z=0.1$, $z=0.2$, $z=0.3,$ and $z=0.4$. The arrowheads indicate the median locations of galaxies in each bin at $z=0$, with their past positions shown at successive redshifts. For statistical reliability, we excluded bins with fewer than five galaxies at $z=0$. For ease of comparison, each plot shows the RS, BC, GV and GZ regions at $z=0$, as defined in Sect. \ref{sec:gv}.

In general, we observe that galaxies classified as blue at $z=0$ experience the greatest growth in stellar mass. In contrast, galaxies classified as green or red at $z=0$, show little to no growth in stellar mass.
Moreover, low-mass red galaxies experience mass loss at low redshift. Green galaxies exhibit the most significant evolution in colour, marking them as ideal candidates for studying the transition from star-forming to quiescence.

\begin{figure}
\centering
{\includegraphics[width=0.4\textwidth]{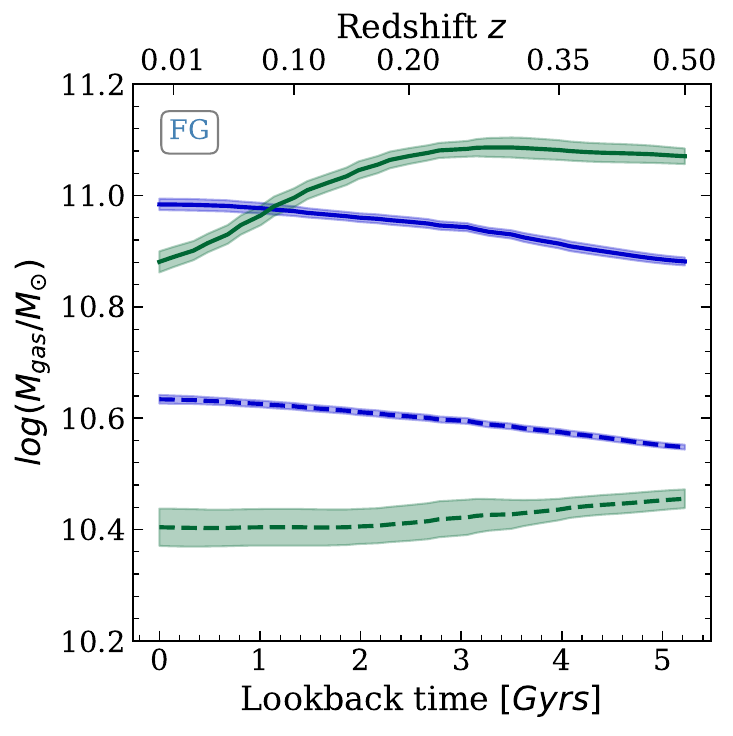}
\caption{\label{fig3b} Median gas mass evolution for blue and green FGs, from $z=0.5$ to $z=0$. Solid and dashed lines represent the trends for high- and low-mass galaxies, respectively, while shaded regions indicate error bars derived through bootstrap re sampling.
}}
\end{figure}

Blue FGs exhibit consistent stellar mass growth and redden over time, regardless of their mass. In contrast, green FGs show a scarcer increase in stellar mass but undergo notable colour changes compared. Red FGs at $z=0$ start, on average, at redder positions in the CMD at $z=0.4$ compared to  blue and green galaxies, evidencing a higher degree of quenching at the starting point of our analysis. There is also evidence of mass loss during the last timestep for red galaxies with $\log(M_\star/M_\odot)\lesssim 10.5$, as a possible cause of stellar evolution: this is typical of ageing stellar populations, where the formation of new stars has significantly decreased or ceased altogether, limiting the ability to offset the losses.

We note that blue and green FGs populations start from similar positions in the CMD at the highest redshift studied. This effect is more pronounced at lower masses. To investigate their evolutionary differences, we examined the possibility that this could be due to: blue galaxies being predominantly centrals, while green galaxies are primarily satellites; green galaxies hosting more active AGNs than their blue counterparts; or blue and green galaxies having different histories of minor and major mergers. However, our analyses suggest that these physical mechanisms are not the primary drivers of the observed differences.  

One possible explanation for why blue and green FGs occupy similar regions in the CMD at $z=0.4$, yet follow markedly different evolutionary trajectories could be related to differences in the evolution of their gas content over time. Figure \ref{fig3b} shows the evolution of gas content in blue and green FGs. Although our primary interest lies in the low-mass regime, we note that for high-mass galaxies, green galaxies had a higher gas content in the past compared to their blue counterparts. However, the key point is that regardless of mass, green galaxies have lost gas over time, whereas blue galaxies have increased their gas content, suggesting a higher inflow of gas. One possible explanation is that galaxies losing their gaseous content may be located near filaments (e.g. \citealt{Hasan:2024, Pandey:2024b}). However, exploring this possibility is beyond the scope of the present work.

Regarding INs, in the first column and second row of Fig. \ref{fig3}, we observe that $z=0,$ blue galaxies exhibit reddening and an increase in stellar mass during the initial time intervals, similar to what is observed in blue FGs. However, during later time intervals, they show a reversal in their colour evolution within the CMD (shifting downward) indicative of a rejuvenation process, which is also accompanied by an increase in stellar mass. The infall of galaxies can trigger temporary increases in star formation, especially in outer regions. This effect is often linked to the specific conditions beyond the virial radius, where the higher galaxy density in a denser environment at the periphery facilitates such bursts (e.g. \citealt{Porter:2008}, \citealt{Mahajan:2012}). For INs that are green or red at $z=0$, the trend is not systematic across the mass range, which complicates a unified analysis. It is plausible, however, that these infalling galaxies arrived as previously isolated centrals or satellites from other systems, potentially experiencing pre-processing, which could contribute to the irregular behaviour observed.

For the other classes studied here (RINs, BSs, and GRs), we observe a similar behaviour within each colour category (blue, green, or red), with evolutionary trajectories in the CMD that are predominantly vertical. This suggests that the physical processes in denser regions primarily drive the notable vertical evolution in the CMD, leading to more significant changes in colour than in the stellar mass.

Blue RINs have, at $z=0.4$, colours that are typically bluer than the ($z=0$) BC ridge line and they reach that line by $z=0$ at the final snapshot. They also show a greater stellar mass evolution compared to their green and red counterparts. Green RINs also start below the BC ridge line at $z=0.4$, but closer to it and show much greater colour evolution than blue RINs. In contrast, low-mass red RINs start redder and closer to the GV, while high-mass red RINs are already classified as green at $z=0.4$, with a behaviour similar to that of red FGs.

During the final time interval analysed, from $z=0.1$ to $z=0$, most RINs cross the system's $R_{200}$. This period we called RIN diving stage. During this period, blue RINs enter the system while located near the BC ridge line, showing little evolution during the RIN phase. In contrast, low-mass green RINs enter the system closer to the GV, while high-mass green RINs enter within the GZ, showing significant colour variation. For red RINs, particularly those with low masses, tidal stripping begins to take effect as they move within the system, leading to notable stellar mass loss and highlighting the environmental impact on the evolution of these low-mass galaxies. Moreover, as in red FGs, stellar evolution also contributes to this mass loss.

For BSs, the BS diving stage corresponds to the time elapsed between their first crossing of $R_{200}$ and their second crossing, representing the duration they spend within the virial region of the system. Following this, the backsplash stage begins when these galaxies exit the virial region, continuing until $z=0$, at which point they can be classified as BSs.

BSs have a median diving time of $1.63 \, \text{Gyr}$, measured as lookback time. This value is consistent with the one reported by \citet{Ruiz:2023} for cluster systems. The similarity in the diving times, despite the different system scales, is explained by the lower velocities of galaxies in groups, which compensate for the smaller spatial scale relative to clusters. In our study, BSs enter the virial region of the system at a median infall time of $z_{in}^{\rm BS} \approx 0.35$, corresponding to a lookback time of $4 \, \text{Gyr}$, and exit from $R_{200}$ at a median time of $z_{out}^{\rm BS} \approx 0.2$, equivalent to a lookback time of $2.5 \, \text{Gyr}$. Thus, the effects of the BS diving time are observed between the first redshift studied ($z=0.4$) and the third ($z=0.2$).

In general, BSs show predominantly vertical movement in the CMD during their backsplash stage (from $z\approx 0.2$ to $z=0$). Blue BSs, compared to blue RINs, experience environmental effects during the BS diving stage such as tidal stripping, galaxy harassment, and ram pressure stripping, which shift them to a higher position in the CMD. For green BSs, we observe a higher median colour than in green RINs. Red BSs, starting from higher CMD positions than green BSs, show deeper evolutionary changes than red RINs. This pattern suggests that either BSs are subject to more intense environmental effects than RINs during the RIN diving stage, or that the effects experienced during the diving phase continue to cause substantial changes in BSs after they leave the system.

\begin{figure*}
\centering
{\includegraphics[width=1\textwidth]{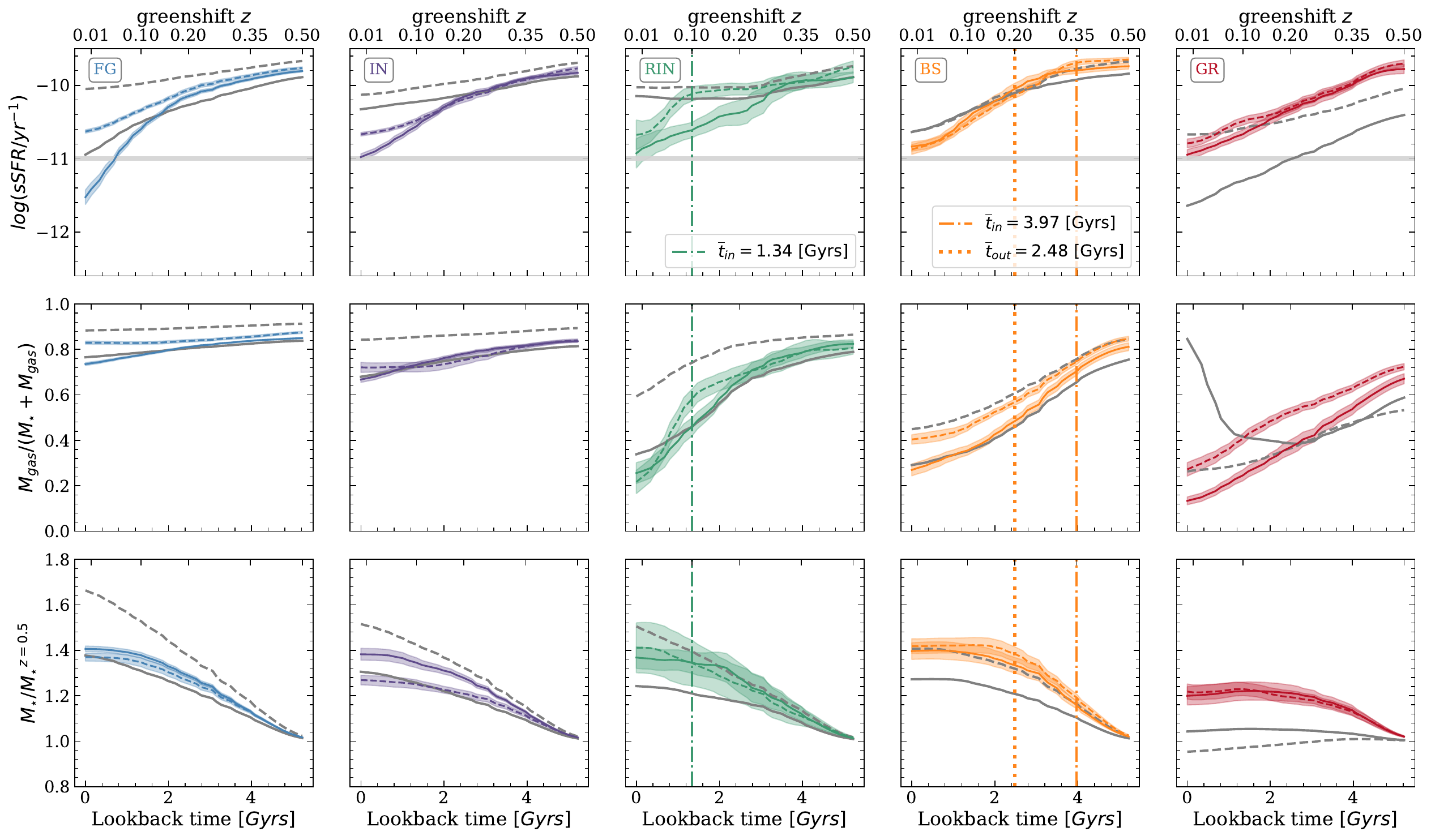}
\caption{\label{fig4} Evolution of green galaxies from $z=0.5$ to $z=0$ for the five classes analysed in this work.  The panels display, from top to bottom: median values of the sSFR, the gas fraction relative to  $M_ {\star}+M_{\rm gas}$, and the stellar mass normalised to its value at $z=0.5$. Solid and dashed lines represent the trends for high- and low-mass galaxies, respectively, separated by the median stellar mass of the random sample. Shaded regions indicate error bars derived through bootstrap resampling. For comparison, we show in grey the behaviour of the total sample for each class, considering not only green galaxies but also blue and red galaxies. In the RIN class panel, the vertical line indicates the median entry time of green RINs into the group's virial region, whereas in the BS class panel, vertical lines mark the median entry and exit times of green BSs from the group’s $R_{200}$. The horizontal light grey line indicates the threshold for classifying galaxies as SF or PS.
}}
\end{figure*}

\begin{figure}
\centering
{\includegraphics[width=0.4\textwidth]{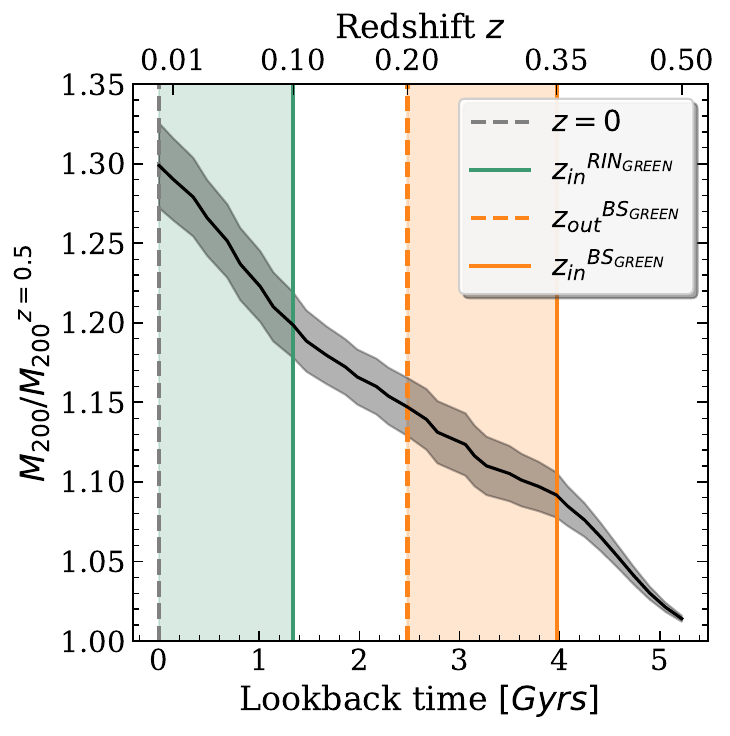}
\caption{\label{fig4b} Median evolution of the $M_{200}$ value for the 214 FoF groups in our sample, normalised to its value at $z=0.5$. The grey shaded regions represent error bars derived from bootstrap resampling. The orange vertical lines indicate the median  ongoing (solid line) and outgoing (dashed line) crossing times of $R_{200}$ for green BSs. The green solid vertical line represents the median group ongoing time for green RINs. The grey dashed line marks $z=0$. The orange shaded region denotes the diving stage for green BSs, while the green shaded area highlights the diving stage for green RINs.
}}
\end{figure}

Finally, it is important to note that, statistically, blue and green GRs are much less frequent than red ones, representing about $8\%$ blue, $12\%$ green, and $80\%$ red. We observe that blue GRs show much smaller colour changes compared to green and red galaxies. To understand why blue GRs experience such minimal evolution, we investigated some of their key characteristics. We analysed the potential bias of these galaxies belonging to less massive FoF groups; however, we did not find any evidence supporting this hypothesis. We find that there are differences in the minimum pericentric distance normalised to the virial radius that these galaxies have during their orbits. In particular, blue GRs have a median pericentric distance normalised to the virial radius of $d_{p} / R_{200} = 0.17$, green GRs have $d_{p} / R_{200} = 0.16,$ and red GRs have $d_{p} / R_{200} = 0.08$. This suggests that red GRs pass much closer to the group centre compared to blue and green ones. The median entry time also shows notable differences among these galaxies: $2.3$ Gyr for blue, $2.2$ Gyr for green, and $4.9$ Gyr for red, suggesting that red GRs have spent significantly more time within the system. Additionally, we observe signs of erosion in red GRs across almost all mass ranges, evidenced by the mass loss of their sub-halos, likely due to tidal effects within the dense group environment. 

%%%%%%%%%%%%%%%%%%%%%%%%%%%%%%%%%%%%%%%%

\section{Transition galaxies}\label{sec:green}

\begin{figure}
\centering
{\includegraphics[width=0.49\textwidth]{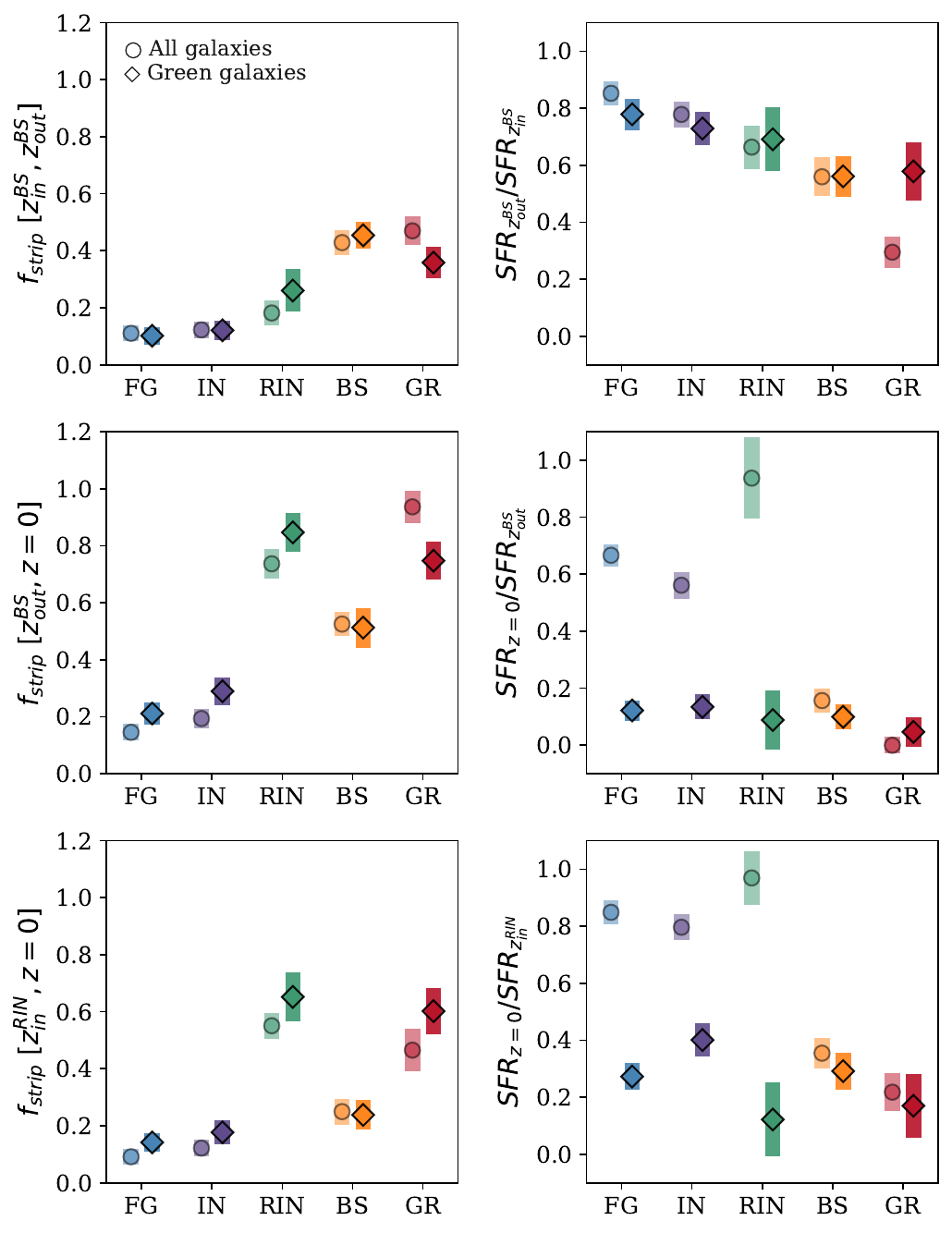}
\caption{\label{fig5} Upper panels: Fraction of mass stripped (left) and SFR ratio (right) during the median diving interval for BSs. Middle panels: Fraction of mass stripped (left) and SFR ratio (right) during the median backsplash time for BSs. Lower panels: Fraction of mass stripped (left) and SFR ratio (right) during the median time interval for RINs within the group. Results are presented for FGs and the four defined types of system galaxies. The values shown represent the median and standard deviation, calculated using the bootstrap method. Circles indicate the total number of galaxies of each type, while diamonds represent only the green galaxies of each type. We recall that in Fig. \ref{fig4}, $z_{in}^{BS} = 0.35$, $z_{out}^{BS} = 0.2$, and $z_{in}^{RIN} = 0.1$.
}}
\end{figure}

\begin{figure}
\centering
{\includegraphics[width=0.45\textwidth]{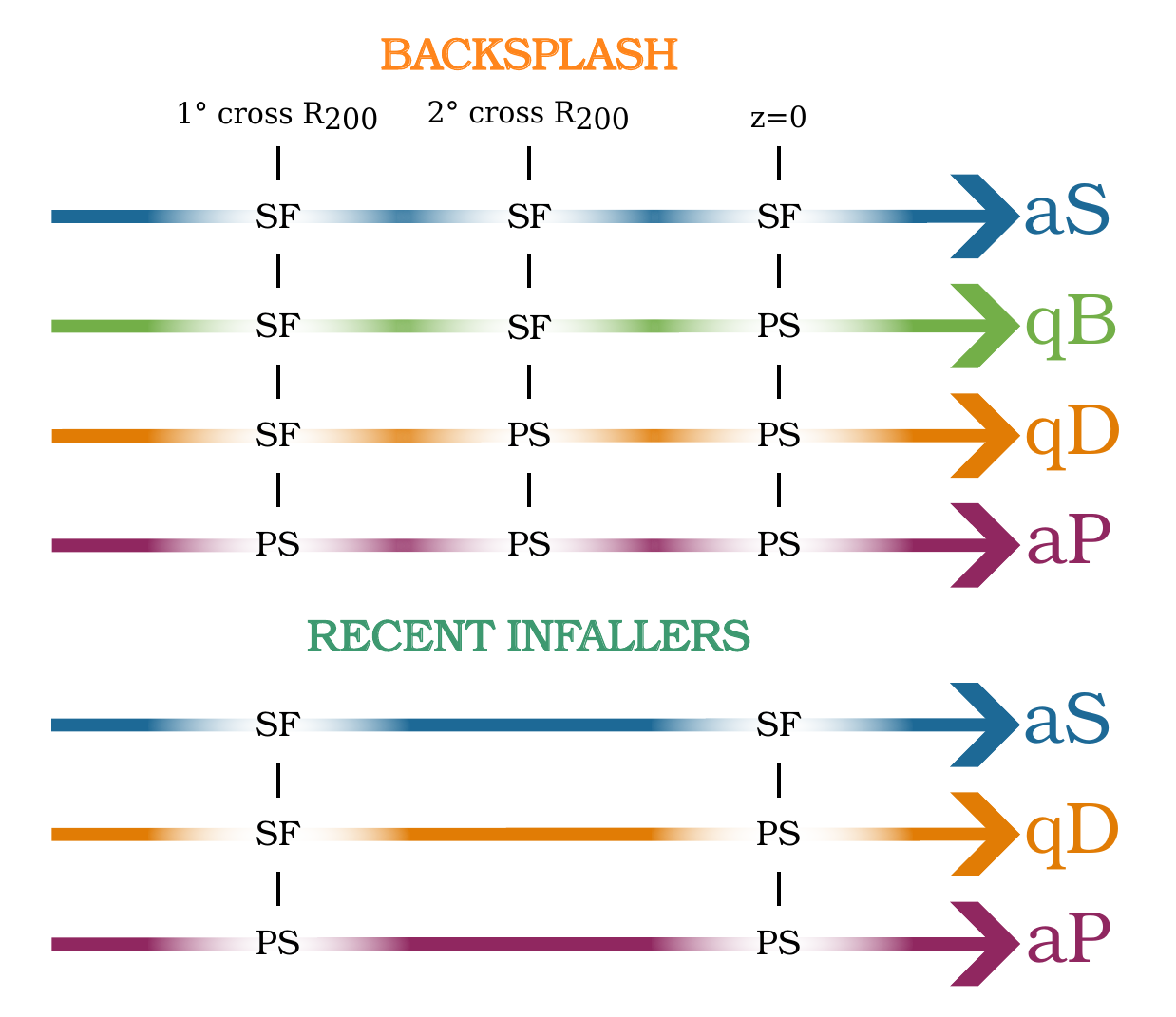}
\caption{\label{fig6} Scheme adopted to classify the sub-types of BSs and RINs is based on the analysis of star formation status at key moments: SF or PS. For BSs, three stages are selected: the moment of entry into the $R_{200}$ region of the group, the moment of exit from that region and at $z=0$. For RINs, two moments are highlighted: the entry into the $R_{200}$ region of the group and $z=0$. The vertical lines delineate the transitions between these stages. The identified sub-types are: always star-forming (aS), quenched in post-splashback stage (qB, only for BSs), quenched during the diving stage (qD), and always passive (aP). Among the green BSs, $44.7\%$ are aS, $39.9\%$ are qB, $15.4\%$ are qD, and none are aP. Among the green RINs, $35.6\%$ are aS, $42.4\%$ are qD, and $22.0\%$ are aP.
}}
\end{figure}

We present evidence in the previous section to support the notion that the most significant evolution in colour occurs INs classified as green at $z=0$. This section examines the evolution of astrophysical properties of green galaxies from $z=0.5$ to $z=0$ across the galaxy classes. We have adopted this redshift range because the primary aim of this study is to examine the physical properties of galaxies during their interaction with the group environment.  Depending on the galaxy, this could correspond to a period before galaxies have interacted with the group for the first time (the diving stage) and in the case of backsplash galaxies, when they are located in the outskirts. The evolution of the different properties is presented in Fig. \ref{fig4}. From top to bottom, the panels display the following properties: the sSFR, the gas fraction relative to $M_{\star}+M_{\rm gas}$, and the stellar mass normalised to its value at $z=0.5$. For each property, we show the median values of the evolution of the stacked galaxy population, with shaded regions representing the error bars derived through bootstrap resampling. As in previous analyses, the samples are divided into high- and low-mass galaxies, with their respective trends represented by solid and dashed lines. For comparison, we show in grey the behaviour of the total sample for each class, without distinguishing by colour, thereby including both blue and red galaxies. In the panels corresponding to the RIN class, the vertical line marks the median entry time of green RINs into the group's virial region. For the BS class panels, vertical lines indicate the median times when these green galaxies entered and exited the group. 

In the IllustrisTNG 300-1 simulations, there is a considerable number of galaxies with $\text{SFR}=0$. These galaxies are excluded from the analysis in Fig. \ref{fig4}. For reference, the percentage of these galaxies at $z=0$, as shown in Table \ref{tab:sfrnull}.

On the other hand, to determine the fractions of star-forming (SF) and passive (PS) galaxies across the different galaxy classes, we analysed the distribution of $\log({\rm sSFR}/\text{yr}^{-1})$ for our environment-independent random sample. A threshold of $\log({\rm sSFR}/\text{yr}^{-1})=-11$ was adopted to distinguish between SF and PS galaxies. In Fig. \ref{fig4}, the horizontal light grey line indicates this threshold. Furthermore, we present the percentages of SF and PS galaxies at $z=0$ for green galaxies across the five classes in Table \ref{tab:sf_ps_green_galaxies}.

As shown in the top row of Fig. \ref{fig4}, all galaxy types exhibit a decrease in their sSFR over time. From $z=0.5$ to $z\sim 0.2$ no mass dependence is observed in any class. However, from $z\sim0.2$ to $z=0$, high-mass green RINs show lower sSFR values compared to their low-mass counterparts. This behaviour is also observed for green FGs, INs and GRs from $z\sim0.1$ to $z=0$. However, in the case of BSs, no dependence on stellar mass is observed. BSs exhibit a further reduction in their sSFR during the diving stage for both low- and high-mass samples, maintaining this declining trend as they transition to the outskirts of the systems. Similarly, green RINs experience a decline in their sSFR (particularly among the low-stellar-mass sample) when they enter the group.

When comparing with the total sample, we observe that for the FGs, INs, and RINs (which at $z=0,$ have a higher percentage of blue galaxies as shown in Fig. \ref{fig2}), the total sample shows a higher sSFR value than the green sample. In the case of BSs, where the percentages of blue, green, and red galaxies are fairly similar, the green sample closely resembles the total sample. Lastly, for GRs, which are predominantly red, the total sample exhibits lower star formation activity compared to the green sample.  

As expected, the highest percentage of passive galaxies in Table \ref{tab:sf_ps_green_galaxies} is found among green GRs. However, we note that the percentage of passive galaxies is higher in RINs green galaxies than in green BSs. From the top panels in Fig. \ref{fig4}, we observe that the impact on galaxies crossing $R_{200}$ since $z\sim 0.1$ is greater than on those that crossed it at earlier times, as seen with BSs.

\begin{table}[ht]
\centering
\caption{Percentage of galaxies with $\text{SFR}=0$ at $z=0$ for all and green galaxies by each class.}
\begin{tabular}{lccccc}
\hline
 & FG & IN & RIN & BS & GR \\
\hline
All galaxies [\%] & 0.0 & 1.7 & 13.5 & 2.8 & 18.0 \\
Green galaxies [\%] & 0.0 & 0.3 & 2.3 & 0.8 & 1.2 \\
\hline
\end{tabular}
\label{tab:sfrnull}
\end{table}

\begin{table}[ht]
\centering
\caption{Percentage of star-forming (SF) and passive (PS) green galaxies for different class of galaxies at $z=0$, with errors obtained using the bootstrap method.}
\begin{tabular}{lccccc}
\hline
Class & FG & IN & RIN & BS & GR \\
\hline
SF [\%] & 47.2 & 47.2 & 37.3 & 44.7 & 32.1 \\
PS [\%] & 52.8 & 52.8 & 62.7 & 55.3 & 67.9 \\
Error [\%] & 1.6 & 2.0 & 6.3 & 3.6 & 3.7 \\
\hline
\end{tabular}
\label{tab:sf_ps_green_galaxies}
\end{table}

\begin{figure*}
\centering
{\includegraphics[width=0.9\textwidth]{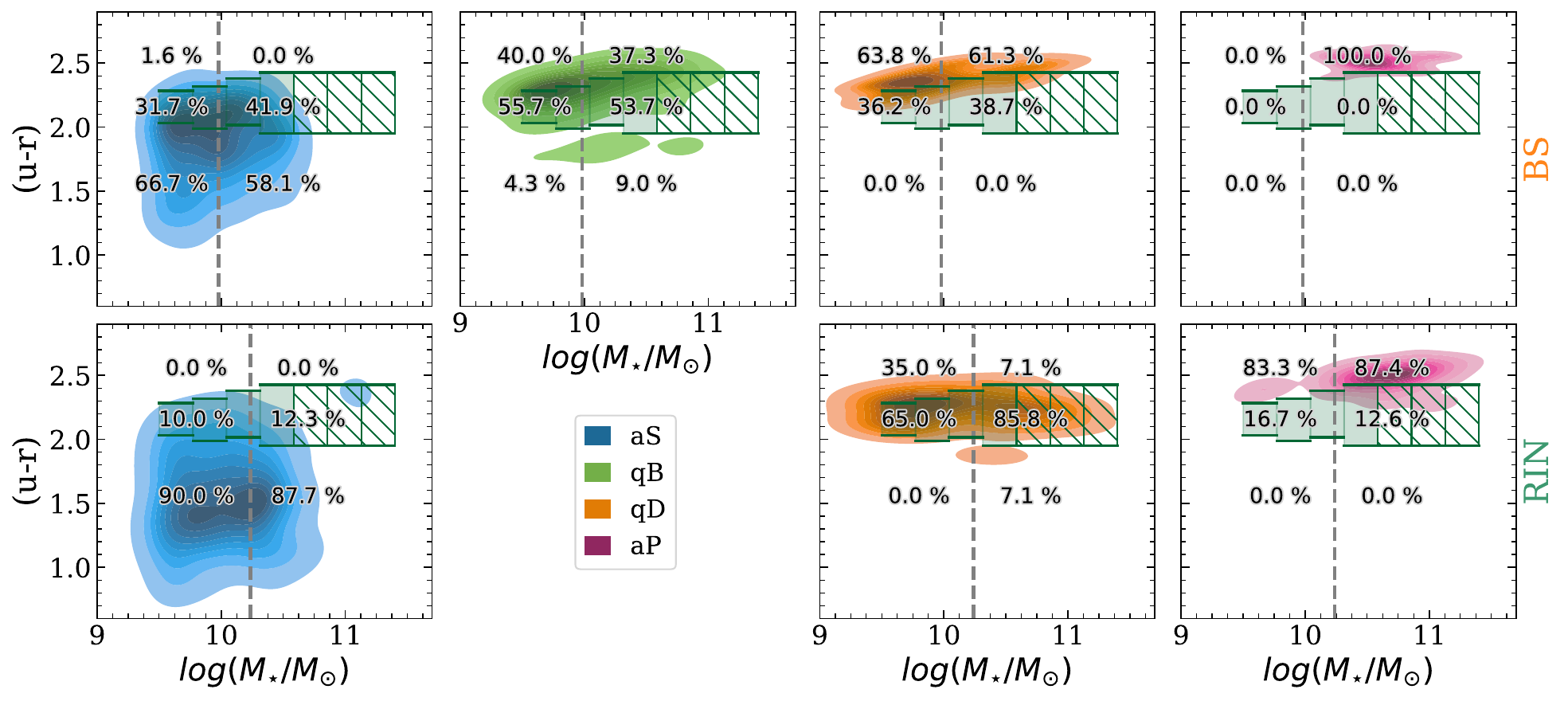}
\caption{\label{fig7} $(u-r)$ colour-stellar mass diagram at $z=0$ for the each BS and RIN sub-types. Upper panels: BS galaxy sub-types are presented from left to right: aS, qB, qD, and aP. The dashed grey vertical line marks the median $\log(M_{\star}/M_{\odot}) = 9.98$  for the BS galaxy sample, dividing the low- and high-mass bins. Lower panels: RIN galaxy sub-types are displayed from left to right: aS, qD, and aP. Here, the dashed grey line corresponds to $\log(M_{\star}/M_{\odot}) = 10.24$. In all panels, colours indicate galaxy number density, ranging from lighter (lower density) to darker (higher density) shades. The shaded green region highlights the GV, while the hatched area marks the GZ as defined in this work. Percentages shown in the plots indicate the fractions of blue (bottom), green (middle), and red (top) galaxies within the low- and high-mass bins. 
}}
\end{figure*}

Therefore, a possible explanation for the observed differences between BSs and RINs lies in the timing of when these two classes first entered the groups; this would make the former suffer the environmental effects of less massive systems than the latter would typically face. This is shown in Fig. \ref{fig4b}, where we present the evolution of $M_{200}$ across redshift. We find that when RINs enter the group, the system’s median $M_{200}$ is approximately 10\% higher than at the time of BS entry. By $z=0$, this difference increases to $20\%$. This suggests that, by the time the RINs enter, the groups already have a larger mass, making it more difficult for the galaxies to maintain ongoing star formation. In other words, more massive groups are more effective at expelling gas from galaxies due to more frequent interactions and stronger gravitational forces, which can lead to gas loss and, consequently, hinder continuous star formation. From an observational perspective, \citet{Vanderburg:2015} found a $20\%$ level of growth in the virial mass of a sample of galaxy clusters between $z=0.25$ and $z=0$. Using a sample of 35 galaxy clusters from the OmegaWINGS survey \citep{Gullieuszik:2015}, \citet{Muriel:2025} found a higher fraction of passive galaxies in the RIN sample compared to the BS sample.

FGs have higher gas fractions compared to those in denser environments. Among the other classes, a progressive decline is observed both over time and across galaxy types, with gas fractions decreasing in the sequence: INs, RINs, BSs, and GRs. Notably, group entry for RINs corresponds to a significant reduction in gas mass, particularly in low-mass galaxies. Compared with the total sample, we find that similarly to the sSFR, FGs, INs, and RINs exhibit a lower gas fraction in the green sample. This effect is particularly pronounced in low-mass galaxies. For BSs, the green sample closely represents the behaviour of the total sample. For GRs, an increase in the gas fraction is observed in the high-mass bin. This is because this population primarily consists of central galaxies within the FoF group sample, and the particles of central galaxies include all particles in the group that have not been associated with any sub-structure by Subfind, namely, intragroup gas and stars. Consistent with this interpretation, the increase in gas fraction is not related to star formation, as it is not accompanied by growth in either the sSFR or stellar mass.

When analysing the growth in stellar mass of galaxies relative to their stellar mass at $z=0.5$, we observe in the bottom row of Fig. \ref{fig4} that the dependence on stellar mass (low or high) during the last 2 Gyr tends to flatten for green galaxies when compared with the corresponding total samples, except for GRs. In green FGs, as well as in green RINs, BSs, and GRs sub-populations, no dependence on stellar mass is detected. However, this dependence is present in green INs, with high-mass green galaxies being the ones that grow the most. It is worth noting that for all classes the total sample shows a dependence on stellar mass in the growth of stellar mass.

To evaluate the impact of the intra-group environment due to ram pressure stripping and strangulation across the various stages considered, we calculated the fraction of gas stripped relative to the total gas available at the end of the considered stages, following \citet{Cora:2018}. This fraction is defined as $f_{strip} = \frac{M_{strip}}{M_{strip} + M_{gas}^{f}}$. Here, $M_{strip}$ represents the total mass of gas stripped during a given stage and is defined as $M_{strip} = \lvert M_{gas}^{f} - M_{gas}^{i} \rvert - \lvert M_{\star}^{f} - M_{\star}^{i} \rvert$, which accounts for the net changes in gas and stellar mass during the stage. We present these results in the left panels of Fig. \ref{fig5}. The top panels correspond to the BS diving stage. The central panels represent the backsplash stage. The bottom panels illustrate the median time from when the RINs enter the system to $z=0$, that is, the RIN diving stage. On the other hand, the right panels show the SFR ratio, defined as ${\rm SFR}^f/{\rm SFR}^i$, for the same stages. The aim is to analyse the effect of these properties on BSs or RINs, depending on the studied interval, in comparison to the other classes at that time. All results are presented for the total sample (circles) and for green galaxies only (diamonds) across the five studied classes.

\begin{figure*}
\centering
{\includegraphics[width=0.7\textwidth]{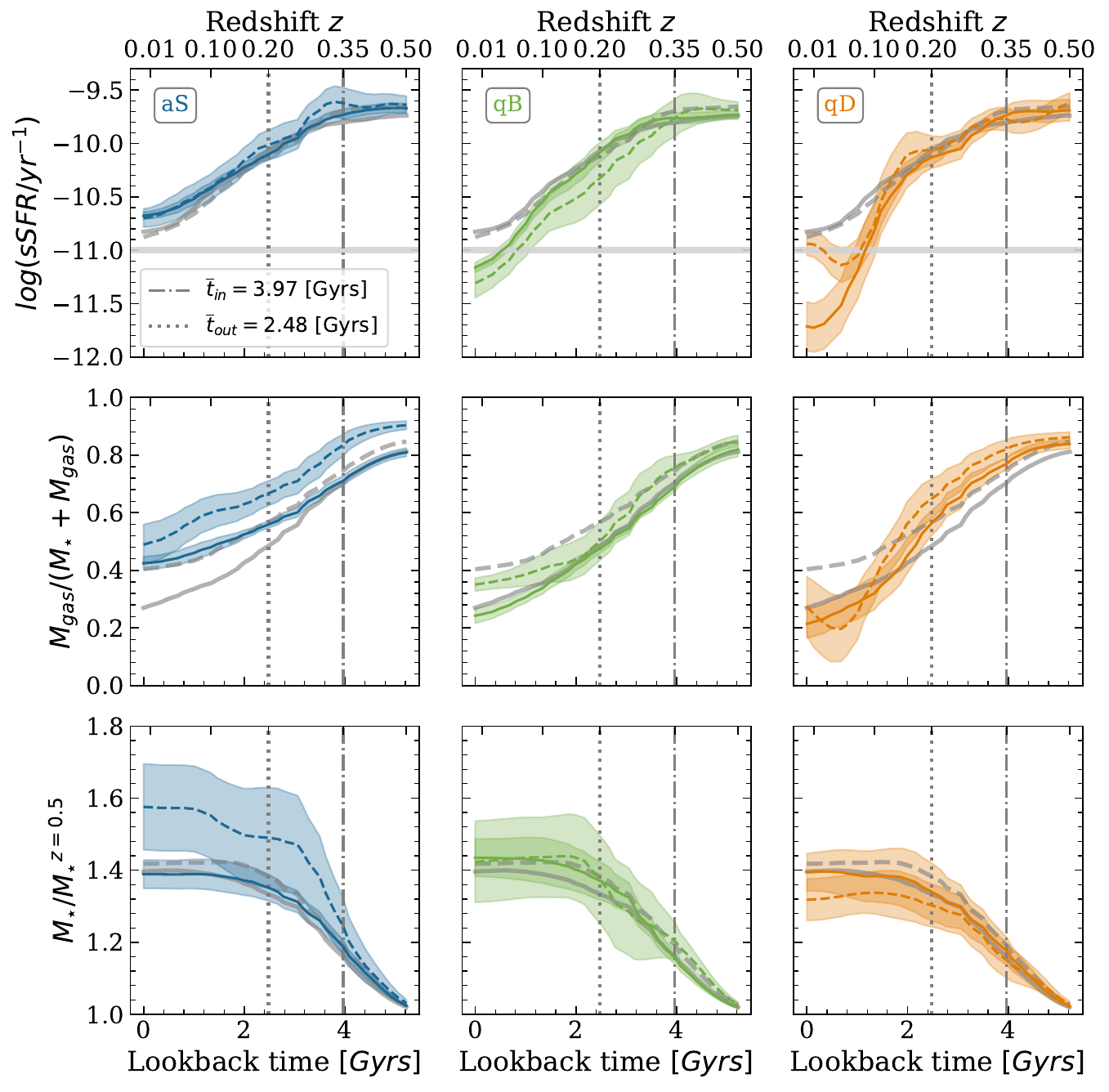}
\caption{\label{fig8} Evolution of green BSs from $z=0.5$ to $z=0$ for the three green BS sub-types. The panels display, from top to bottom: the sSFR, the gas fraction relative to  $M_ {\star}+M_{\rm gas}$, and the stellar mass normalised to its value at $z=0.5$. Solid and dashed lines represent trends for high- and low-mass galaxies, respectively, with shaded regions indicating bootstrap-derived error bars. For comparison, the results found in Fig. \ref{fig4} for the total sample of green BSs are reproduced across all properties. The horizontal light grey line indicates the threshold for classifying galaxies as SF or PS. The vertical lines mark the median entry and exit times of green BSs from the group’s $R_{200}$.
}}
\end{figure*}

The fraction of stripped baryonic mass provides insight into galaxies' ability to lose gas and stellar mass over a given time interval, while the ratio of SFR reflects their ability to form new stars during the same period. It is important to note that galaxies that start this interval with $M_{\text{gas}}^{i} = 0$ are not included in this count. Examining the diving stage for BSs across the different classes (upper panel of Fig. \ref{fig5}), we find that BSs experience the highest stripping in this stage. GRs also show high levels of stripping, with GV galaxies being slightly less stripped compared to the entire sample. However, this difference is within the error margins. For RINs, which are predominantly blue, the stripping is lower than that of the BS and slightly more pronounced in the green population compared to the overall RINs population. The SFR ratio during the diving stage of BS follows a consistent trend: FGs exhibit the highest star-forming capacity, followed by IN and RINs with similar values, while BS and GRs have lower SFR ratios, both in the total sample and among green galaxies. A clear correlation exists between the degree of baryonic mass loss and the decline in SFR during the diving stage for BSs. For GRs, we note that the decrease in the SFR is significantly greater in the total sample than in the green galaxies, consistent with the fact that GRs are predominantly red.

During the backsplash stage, BSs experience significant stripping; however, their mass loss is less pronounced compared to GRs at this stage (regardless of their colour), with green RINs undergoing the greatest mass loss. During the same period, INs maintained low stripping levels, comparable to those of FGs. The SFR ratio during this phase indicates that all classes of galaxy undergo a reduction in the star-forming capacity. However, for FGs, INs, and RINs, there is a marked difference between the total sample per class and the green galaxies, with the latter showing a greater decline in star formation.

In the final interval studied ($z_{in}^{RIN}, z=0$, lower panel of Fig. \ref{fig5}), corresponding to the RIN diving phase, a
pattern similar to the previous interval is observed, likely due to their comparable timescale. However, mass stripping increases significantly for RINs, particularly for the green sub-sample, reaching levels that are higher than those observed in other types. The SFR ratio continues to follow a consistent trend, closely resembling the previous interval.

\begin{figure*}
\centering
{\includegraphics[width=0.8\textwidth]{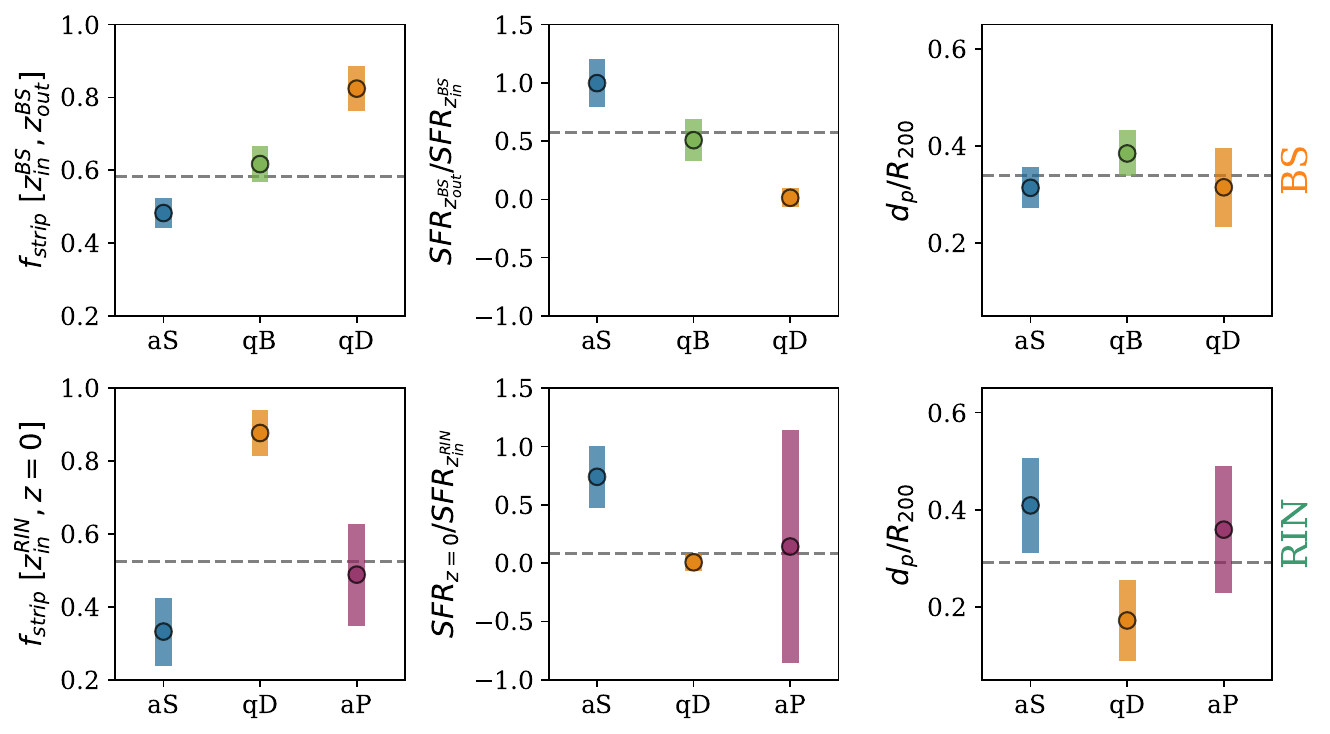}
\caption{\label{fig9} Fraction of stripped mass (left), SFR ratio (centre), and minimum pericentric distance normalised to the virial radius (right) for the green BS galaxy sub-types (upper panels) are shown for the period from their entry into the $R_{200}$ region until they exit. Similarly, for the green RINs sub-types (lower panels), the analysis covers the period from their entry into the $R_{200}$ region until $z=0$. Circles represent the median values, while bars indicate the standard deviation, calculated using the bootstrap method for each sub-type. The dashed grey horizontal lines denote the median values of the green galaxies of the corresponding properties.
}}
\end{figure*}

\subsection{Backsplash galaxies}\label{sec:BS}

Given that the highest proportion of green galaxies come from the BS class, this section examines the evolution of green BSs and their relationship to their passage through the group.
We classified four sub-types within BSs. Following \citet{Ruiz:2023}, we categorised these galaxies based on whether they are SF or PS at the time of their first crossing through $R_{200}$, at their second crossing through $R_{200}$, and at $z=0$. We find that the median infall time for green BSs occurred $4$ Gyr ago, with their median exit time being $2.5$ Gyr ago. We use the same threshold specified in Sect. \ref{sec:green} to distinguish between SF and PS galaxies. Galaxies that remain SF throughout all stages are classified as always star-forming (aS). Those that quench at some point between exiting the group and $z=0$ are classified as quenched at the backsplash stage (qB). Galaxies that cease star formation while within the group are labelled as quenched at diving stage (qD). Finally, those that are passive throughout all stages are categorised as always passive (aP). The upper part of Fig. \ref{fig6} provides a schematic representation of the four sub-types of BSs. 

The upper panels of Fig. \ref{fig7} present the isodensity contours in the $(u-r)$ colour-stellar mass diagram, focussing on the different sub-classes of BSs. We observe a decrease in the fraction of galaxies in the BC and the corresponding increase in the fraction of galaxies in the RS from the aS to the aP sub-type. The highest proportion of green galaxies is found in the qB sub-type, with approximately half of the galaxies in both low- and high-mass ranges occupying this region. Additionally, BSs in the aS and qD sub-types each represent between $30\%$ and $40\%$ of their population across both mass ranges within the GV and GZ regions. 

In Fig. \ref{fig8}, we analyse the same quantities as in Fig. \ref{fig4}, but specifically for the green BS subclasses of galaxies. For comparison, we include the trends for the total green sample of BSs in grey. Regarding the sSFR, we observe that mass dependence is generally minimal throughout the interval, except in the case of qD galaxies at lower redshift. The greatest decline in specific star formation occurs in the qD sample. Given our threshold of $\log({\rm sSFR}/\text{yr}^{-1}) = -11$ to distinguish star-forming galaxies from passive ones, it is notable that up to about 1 Gyr ago, both low- and high-mass qD galaxies fall below this threshold. This behaviour is accompanied by a greater reduction in gas fraction compared to the other sub-classes. Beyond this behaviour, it is worth noting that, on average, low-mass qD galaxies show a tendency to rejuvenate in terms of star formation. This could be linked to star formation processes induced by ram pressure, as observed in Jellyfish galaxies (e.g. \citealt{Poggianti:2019, Lee:2022, Roberts:2022}).

Lastly, in terms of stellar mass growth relative to initial mass, low-mass aS galaxies show the most significant growth due to their ongoing star-forming activity in this BS phase.
When comparing among the green BSs sub-types and the total green BS sample, we observed that the  aS and qB classes, which represent the largest percentages of green BSs, best reflect the behaviour of the total green BS sample throughout the studied time interval.

\begin{figure*}
\centering
{\includegraphics[width=0.7\textwidth]{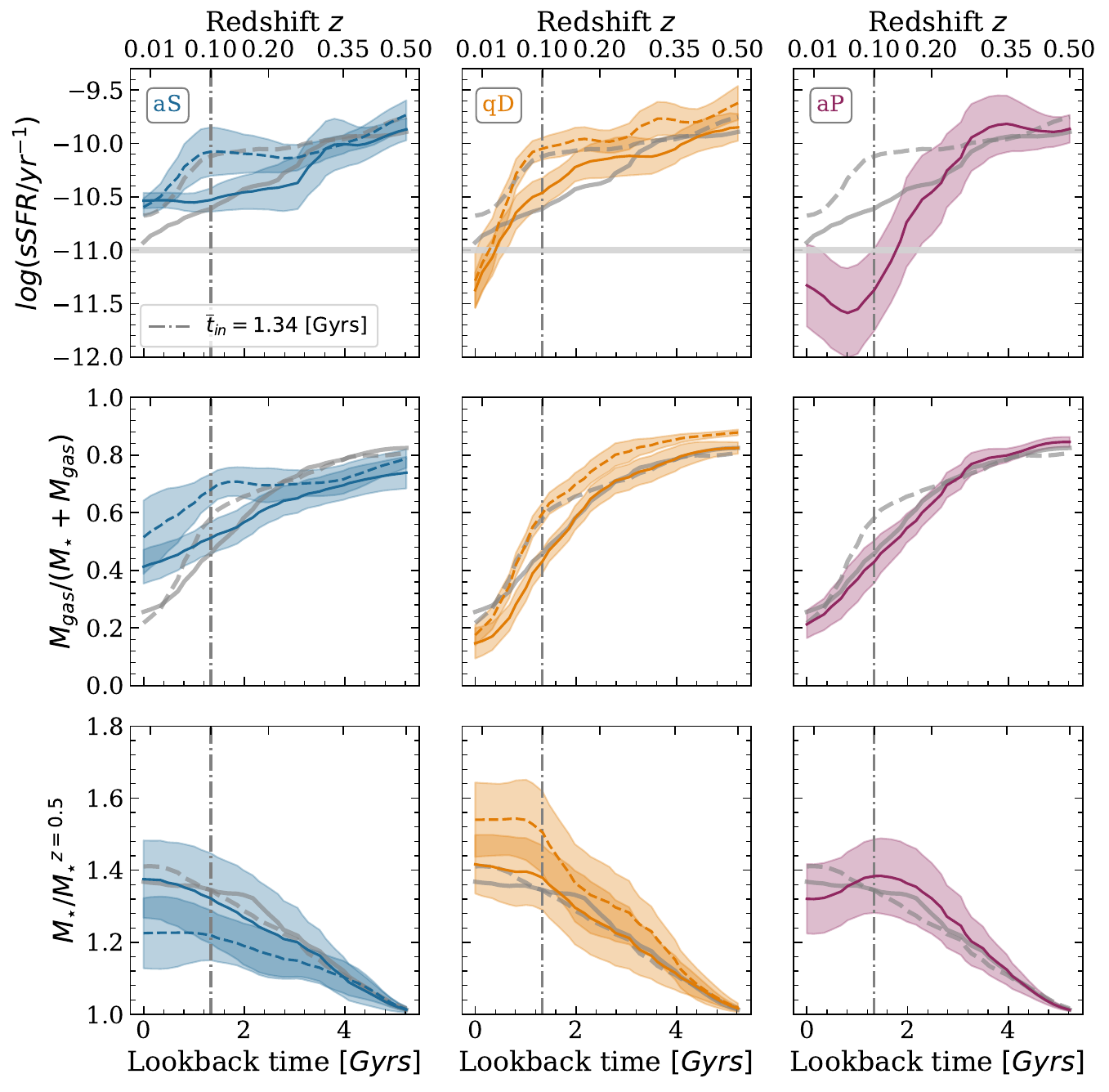}
\caption{\label{fig10}  Evolution of astrophysical properties of green RINs from $z=0.5$ to $z=0$ for the three sub-classes. The panels display, from top to bottom: Specific SFR,  gas fraction relative to  $M_ {\star}+M_{\rm gas}$, and  stellar mass normalised to its value at $z=0.5$. Solid and dashed lines represent trends for high- and low-mass galaxies, respectively, with shaded regions indicating bootstrap-derived error bars. For comparison, the results found in Fig. \ref{fig4} for the total sample of green RINs are reproduced across all properties. The horizontal light grey line indicates the threshold for classifying galaxies as SF or PS. The vertical line indicates the median entry time of green RINs into the group's virial region.
}}
\end{figure*}

In the upper panels of Fig. \ref{fig9}, we analyse the stripped mass fraction, the ratio of SFR during the median time of the BSs diving phase, and the median of the minimum pericentric distance normalised to the virial radius for the three sub-types of green BSs: aS, qB, and qD. We observe that the stripped baryonic mass fraction is highest in qD galaxies, which show the largest decrease in star formation.  In contrast, aS galaxies experience less stripping and maintain an SFR ratio close to one (indicating little to no loss in star formation activity). The qB galaxies show intermediate values for both stripped mass fraction and SFR ratio between the aS and qD types. The average pericenter distances of the different classes are statistically similar. The reference trends for the total green BS sample are similar to those of the qB BSs.

\subsection{Recent infallers}\label{sec:RIN}

Given that green RINs exhibit a higher fraction of passive galaxies at $z=0$ compared to BSs (see Table \ref{tab:sf_ps_green_galaxies}), we analysed the evolution of their properties to understand the differences between these two galaxy classes. The lower section of Fig. \ref{fig6} presents a schematic diagram illustrating the three sub-types of RINs, which are defined by their star formation state. They are basically the same sub-types we defined for BSs with the obvious exception of the qB sub-type since there is no backsplash stage for RINs.

Firstly, we analyse the distribution of blue, green, and red galaxies for each RIN sub-type, distinguishing between high- and low- stellar mass galaxies. The lower panels of Fig. \ref{fig7} show the number density in the CMD for the aS, qD, and aP sub-types (from left to right). Similarly to the BSs, and as expected, we observe a gradual transition from a higher density in the BC for aS galaxies to a predominance in the RS for aP galaxies, with qD being the intermediate case. Among the RIN sub-types, qD galaxies exhibit the highest percentage of green galaxies.

In Fig. \ref{fig10}, we show the evolution of astrophysical properties for the green RINs sub-types, following an approach analogous to that presented in Fig. \ref{fig8}. We find that the median infall time for green RINs was only 1.33 Gyr, in contrast to the 4 Gyr that have passed since BSs first entered the virial region of the groups, with a median exit time of 2.5 Gyr ago, as mentioned in the previous section. For aP galaxies, the analysis is restricted to the high-mass population for statistical reasons. The grey reference lines represent the behaviour of the total green RIN galaxy sample for each property analysed.

For all sub-types, we observed a decline in the star formation activity as galaxies lose their gas content. In the case of low-mass aS galaxies, the impact is more pronounced after they enter the group. A similar trend is observed in qD galaxies, regardless of mass, where the group’s influence on star formation and gas fraction is particularly strong. In contrast, high-mass aP galaxies exhibit reduced star formation and gas loss much earlier, suggesting that they are predominantly affected by pre-processing mechanisms before reaching the group's virial region.

Regarding the stellar mass normalised to the initial mass over the time interval studied, we observe that after the entry time of RINs into the group, the aS and qD sub-types exhibit a stagnation in their stellar mass growth. In contrast, the aP galaxies show a decrease in stellar mass, which could be attributed to tidal stripping.

Finally, as we did for green BSs,  we investigated the stripped mass fraction, the SFR ratio, and the minimum pericentric distance normalised to the virial radius reached from the time they enter the group until $z=0$  for the green RINs (as shown in the lower panels of Fig. \ref{fig9}). It is clear that qD galaxies are much more stripped, exhibit a decrease in star formation, and pass much closer to the group centre compared to aS galaxies. This is in good agreement with the results shown in Fig. \ref{fig10}, where we observe differences in sSFR and gas fraction after entering the group. On the other hand, although aP galaxies also show characteristics of being stripped, the high dispersion, primarily caused by the low number of galaxies, prevents us from making further inferences about the behaviour of these galaxies.

%%%%%%%%%%%%%%%%%%%%%%%%%%%%%%%%%%%%%%%%

\section{Conclusions}\label{sec:conclusions}

This work analyses the evolution of galaxies in different environments, focussing on their transition through the GV and the GZ in intermediate-mass galaxy groups ($13.5 \leq \log(M_{200}/M_{\odot}) \leq 13.7$). Using the Illustris TNG300-1 hydrodynamical cosmological simulations, we studied the trajectories of galaxies in our sample to define five distinct classes: group galaxies (GRs), backsplash galaxies (BSs), galaxies that have recently fallen into a group (RINs), infalling galaxies (INs), and galaxies in the field (FGs).  

Galaxies in these environments are classified as blue, green, or red based on their optical colour in the $(u-r)$-stellar mass diagram (CMD). All galaxies in our study have $\log(M_{\star}/M_{\odot}) \geq 9.5$. Special emphasis is placed on the BS and RIN classes to elucidate the impact of both internal and external quenching mechanisms on galaxy evolution. Our findings highlight several key points:

\begin{itemize}
    \item At $z=0$, FGs predominantly occupy the blue cloud (BC), while those in groups and their surroundings exhibit progressive reddening in the sequence: INs, RINs, BSs, and GRs, particularly among low-mass galaxies, driven by environmental effects. In the group environment, IN and RINs dominate the BC, reflecting active star formation; whereas BSs show the highest fraction of green galaxies. This suggests  their initial dive into groups facilitates the transition to quiescence. In contrast, GRs are primarily found on the red sequence (RS).

    \item  From $z=0.4$ to the present, FGs, INs, and RINs that are blue at $z=0$ exhibit the greatest stellar mass growth with mild colour evolution. Notably, INs display signs of rejuvenation, suggesting temporary increase in star formation, likely triggered by pre-processing events, similar to those observed in jellyfish galaxies. While BSs that are blue at $z=0$ also grow in stellar mass, they do so to a lesser extent than the previously mentioned classes and show greater reddening, especially during the last time intervals considered. Notably, blue GRs are rare ($\sim8\%$ of all GRs) and experience minimal colour evolution compared to green and red ones. We find that this is related to their orbital histories: blue GRs have a median pericentric distance of $ d_{p} / R_{200} = 0.17 $, larger than that of green ($ d_{p} / R_{200} = 0.16 $) and red ($ d_{p} / R_{200} = 0.08 $) GRs. Additionally, they have the shortest median entry times ($2.3$ Gyr), compared to $2.2$ Gyr for green and $4.9$ Gyr for red GRs, meaning they have spent less time within the group environment. These results suggest that their larger pericentric distances and shorter residence times reduce their exposure to the dense group environment, limiting their colour evolution. Conversely, red GRs, which have spent significantly more time within the group and typically pass closer to its centre, show signs of erosion across almost all mass ranges, as evidenced by sub-halo mass loss likely driven by tidal effects.
    
    \item  Galaxies classified as green at $z=0$ undergo the most pronounced colour changes since $z=0.4$, regardless of their classification. For low-mass galaxies, these changes are primarily driven by environmental effects, such as ram pressure stripping and strangulation \citep{Nelson:2018}. Among green galaxies, BS exhibit the most significant colour changes.
    
    \item In general, low-mass galaxies that are red at $z=0$ exhibit a decline in stellar mass, likely driven by environmental processes such as tidal stripping. Red BSs show the most significant colour evolution, showing that a single passage through the system can drive a galaxy from its star-forming phase to quiescence.
\end{itemize}

We examined some of the properties of green galaxies, including the stellar mass, sSFR, and gas fraction, as well as their evolution over time. Our main findings can be summarised as follows.

\begin{itemize}
    \item A general decline in the sSFR over time is observed across all green galaxy classes. Low-mass green RINs exhibit a different behaviour compared to the rest, maintaining a high SFR until they enter $R_{200}$. After this point, they show a sharp decline in sSFR, suggesting that during the infall stage, these galaxies might have experienced increased star formation due to phenomena similar to those seen in jellyfish galaxies, followed by a rapid quenching.
    
    \item FGs exhibit the highest gas fractions at $z=0$, while galaxies in groups and in their surroundings show a progressive decline in gas content, with a more pronounced drop in RIN and GRs. This is consistent with the observation that green RINs have a higher fraction of passive systems compared to green BSs. These findings suggest that the timing of group entry is a factor to reckon with in a galaxy's evolution towards passive states. GRs entered the group earlier and have been exposed to the system's environmental effects for a longer period, whereas RINs entered a system where environmental effects are stronger than those experienced by BSs at the time of their entry.

    \item During the diving phase of BSs, green galaxies experience the most significant gas loss. In contrast, GRs, while also exhibiting substantial losses, they show a lower gas content at the beginning of this phase. During the backsplash phase, BSs do not show significant mass loss, whereas GRs and RINs undergo notable losses over the same period of time. 
    
    \item In general, the stellar mass of green galaxies increases over time from $z=0.5$ to $z=0$, with a clear plateau during the last 2 Gyrs. This behaviour is independent of the stellar mass of the galaxies at $z=0$, with the sole exception of INs, which show greater growth for high-mass galaxies.

\end{itemize}

Given the higher proportion of green galaxies among BSs systems and the greater fraction of passive green galaxies at $z=0$ in RINs systems, we examined these two classes in greater detail. BSs are sub-divided into four types, while RINs are categorised into three types, based on their star-forming or passive characteristics. Our main findings are:

\begin{itemize}
    \item BSs quenched during the backsplash stage constitute approximately half of the green galaxy population across both high- and low-mass systems, predominantly residing in the GV.
    \item In terms of sSFR, BSs quenched during the diving time exhibit the most significant decline in star formation. This behaviour is due to a lower gas reserve compared to other sub-classes. In addition, some low-mass galaxies quenched during the diving time show signs of rejuvenation, with an increase in the star formation that could be linked to ram-pressure-induced star formation processes, similar to those observed in jellyfish galaxies.
    \item No statistically significant differences are observed in the minimum pericentric distance normalised to the virial radius among the different BS sub-classes.
    \item RINs quenched during the diving time have the highest percentage of green galaxies.
\end{itemize}

Regarding the characteristic timescales, BSs crossed $R_{200}$ approximately 4 Gyr ago and left it about 2.5 Gyr ago, with a median diving phase spanning 1.63 Gyr. In contrast, RINs entered the group more recently (1.34 Gyr ago) and show more pronounced evolutionary changes over shorter timescales compared to BSs. This analysis suggests that the timing of entry into the group has an important impact on their evolution. BSs, by entering earlier, do so into a less massive system than RINs. This causes the latter to experience greater influence from the intragroup medium compared to BSs.

Our results show that infall time and environmental effects, such as gas depletion, are closely linked to the decline in sSFR and the transition through the GV towards a passive state. This is particularly evident in RINs and BSs, which exhibit greater mass loss and reduced star formation as they enter and settle in denser regions.

Our study is based on cosmological simulations, providing a theoretical perspective on galaxy evolution in groups. A natural next step is to explore how these trends manifest in observational data. In a future work, we plan to analyse projected quantities in simulations and compare them with observations to assess the extent to which these theoretical predictions hold in real systems.

%%%%%%%%%%%%%%%%%%%%%%%%%%%%%%%%%%%%%%%%

\begin{acknowledgements}
We gratefully acknowledge financial support from the Argentinian institutions: Consejo Nacional de Investigaciones Cient\'ificas y T\'ecnicas (CONICET; PIP-2022-11220210100064CO), from the Agencia Nacional de Promoci\'on de la Investigaci\'on, el Desarrollo Tecnol\'ogico y la Innovaci\'on de la Rep\'ublica Argentina (PICT-2020-3690), from the
Secretar\'ia de Ciencia y Tecnolog\'ia de la Universidad Nacional de C\'ordoba (SECYT-UNC, Res. 258/53).

MdlR is supported by the Next Generation EU program, in the context of the National Recovery and Resilience Plan, Investment PE1 – Project FAIR ``Future Artificial Intelligence Research'', the Comunidad Aut\'onoma de Madrid and Universidad Aut\'onoma de Madrid under grant SI2/PBG/2020-00005, and by the Spanish Agencia Estatal de Investigaci\'on through the grants PID2021-125331NB-I00 and CEX2020-001007-S, funded by MCIN/AEI/10.13039/501100011033.

The TNG 300-1 simulations used in this work are part of the IllustrisTNG project, which were run on the HazelHen Cray XC40 system at the High-Performance Computing Center Stuttgart, as part of the GCS-ILLU project of the Gauss Centers for Supercomputing (GCS).
\end{acknowledgements}

%%%%%%%%%%%%%%%%%%%%%%%%%%%%%%%%%%%%%%%%

\bibliographystyle{aa} 
% \bibliography{biblio} 

\end{document}